\documentclass[12pt]{article}
\usepackage{tipa}
\usepackage{amssymb}
\usepackage{amsfonts}
\usepackage{color}
\usepackage{footmisc}
\usepackage{ulem}
\usepackage[numbers,sort&compress]{natbib}
\usepackage[bookmarksopen=true, bookmarksopenlevel=\maxdimen,
bookmarksnumbered=true,colorlinks,linkcolor=blue,
citecolor=blue,urlcolor=blue]{hyperref}

\newcommand{\omits}[1]{}

\definecolor{dyellow}{rgb}{1.,0.8,.0}
\definecolor{myblue}{rgb}{.1,.1,.7}
\definecolor{dcyan}{rgb}{.0,.6,.6}
\definecolor{dmagenta}{rgb}{0.6,0.0,0.6}
\definecolor{brown}{rgb}{0.6,0.2,0.}
\definecolor{darkblue}{rgb}{.0,.0,0.5}
\definecolor{darkred}{rgb}{0.75,0.0,0.0}
\definecolor{orange}{rgb}{1.,.6,.0}
\definecolor{dorange}{rgb}{0.8,.4,.0}
\definecolor{darkgreen}{rgb}{0.0,0.6,0.0}
\definecolor{purple}{rgb}{.4,.0,.4}
\definecolor{lightgrey}{rgb}{0.7,0.7,0.7}

\begin{document}
\phantomsection \addcontentsline{toc}{chapter}{Weak field
approximation in a model of de Sitter gravity: Schwarzschild-de
Sitter solutions}
\begin{center}
{\bf \LARGE Weak field approximation\\
in a model of de Sitter gravity:\bigskip\\
Schwarzschild-de Sitter solutions}

\bigskip

\bigskip

{\Large Jia-An Lu$^{a,b}$\footnote{Email: ljagdgz@163.com} and
Chao-Guang Huang$^{a}$\footnote{Email: huangcg@ihep.ac.cn}}

\bigskip

$^a$ Institute of High Energy Physics, and
Theoretical Physics Center for \\
Science Facilities, Chinese Academy of Sciences, Beijing 100049,
China

$^b$ School of Mathematics and Computational Science, Sun Yat-sen
University, Guangzhou 510275, China

\phantomsection \addcontentsline{toc}{section}{Abstract}
\begin{abstract}
The weak field approximation in a model of de Sitter gravity is
investigated in the static and spherically symmetric case, under the
assumption that the vacuum spacetime without perturbations from
matter fields is a torsion-free de Sitter spacetime. It is shown on
one hand that any solution should be singular at the center of the
matter field, if the exterior is described by a Schwarzschild-de
Sitter spacetime and is smoothly connected to the interior. On the
other, all the regular solutions are obtained, which might be used
to explain the galactic rotation curves without involving dark
matter.
\end{abstract}
\end{center}

\section{Introduction}
Many works have been done on the gauge theories of gravity
\cite{Hehl76}, motivated by the similarities between Einstein's
general relativity (GR) and Yang-Mills gauge theory. A variety of
models with different dynamics have been studied, among which there
is a model of de Sitter (dS) gravity \cite{Wu74,An76,Guo79,dS2,dS3}
with a gauge-like action constructed by a dS algebra-valued
connection. The Einstein term and a cosmological term can be deduced
from the gauge-like action besides for two quadratic terms of the
curvature and torsion. If the coefficient of the Einstein term is
required to be much lager than that of the quadratic curvature term,
the cosmological constant deduced from the gauge-like action should
be very large. The large cosmological constant may be canceled out
by the vacuum energy density, leaving a small cosmological constant
\cite{dS2}. But it is difficult to explain why the large
cosmological constant and the vacuum energy density are so close,
but not exactly equal, to each other. On the other hand, if the
cosmological constant deduced from the gauge-like action is required
to be small \cite{Wu74,An76,Guo79,dS3}, the coefficient of the
Einstein term should be much smaller than that of the quadratic
curvature term. The model under this case, hereafter referred to as
the dS gravity model or simply as the model, might be very different
from GR. But fortunately, it has been shown \cite{vacuum1,vacuum2}
that all torsion-free vacuum solutions of the model are the vacuum
solutions of GR with the same cosmological constant, and vise versa.
Moreover, it has been shown \cite{{cos1},{cos2}} that the model may
explain the accelerating expansion of the universe and supply a
natural transit from decelerating expansion to accelerating
expansion without the help of dark energy. Of course, further
investigations are needed to check whether the model could explain
various experimental observations consistently.

In GR with a cosmological constant, the Schwarzschild-de Sitter
solution (S-dS) \cite{Kottler}, instead of the Schwarzschild
solution in GR without cosmological constant, is used to explain the
solar-system-scale observations, (such as, the precession of Mercury
perihelion, light deflection, gravitational redshift, radar echo,)
when the cosmological constant is small enough \cite{Solar-SdS}.
Although the vacuum Schwarzschild or the vacuum S-dS solution is
sufficient and the details of the internal solution are not
necessary for the explanation of these tests, a gravitational
theory, as a fundamental theory, should permit regular, reasonable
sources to generate such a kind of the field.  In other words, a
Schwarzschild or S-dS solution should be able to join with regular and reasonable
internal solution under suitable junction conditions.  It is well
known that in GR there exist the regular internal solutions joining
with the Schwarzschild or S-dS exterior.  Moreover, any alternative,
reasonable gravitational theory should be able to reduce to the
Newtonian law of gravity in the weak field approximation in order to
explain the daily-life gravitational phenomena, which also requires
that the vacuum Schwarzschild or the vacuum S-dS solution can be
connected to the internal solution with a non-uniform distribution
of matter.

The S-dS spacetime, as one of the vacuum solutions of the dS gravity
model, is also important for the model to explain the observations
within the solar system scale, for the test particles without spin
current, including light, move along the same geodesics in the S-dS
spacetime obtained in different theories \cite{Zhang73,An76,Hehl76}.
However, the existence of regular internal solution with the smooth
junction to the S-dS solution is still not clear in the dS gravity
model. It has been shown that the energy-momentum-stress tensor of a
spinless fluid should be with constant trace in the torsion-free
case of the model \cite{cos1}. For the general case, the trace of
the energy-momentum-stress tensor is not a constant. It shows that
the torsion-free condition is not so reasonable in the general case.
This property together with the smooth junction condition may
stimulate the search for S-dS solutions with nonzero torsion.
Although the different dS spacetimes with nonzero torsion in the
model have been obtained in Refs. \cite{dStor1,dStor2}, they are
still not the S-dS solutions. For many quadratic models of gauge
theory of gravity, S-dS solutions with spherically symmetric torsion
have been given \cite{S-dS}, but our model does not fall into those
cases.

One may firstly check the existence of the smooth junction in the
weak field approximation. For GR with a cosmological constant,
it is easy to shown that there exist the regular internal solutions
with the smooth junction to the S-dS exterior, in the static,
spherically symmetric weak field approximation. But for the dS
gravity model, more efforts should be taken on the weak field
approximation.  The Newtonian limit of the general quadratic models
without cosmological term has been analyzed \cite{{NL1},{NL2}} in
the 1980's. In those papers, the vacuum spacetime without
perturbations from matter fields is the Minkowski spacetime. But in
the theory with a cosmological term, the vacuum spacetime without
perturbations should be a dS spacetime. The perturbations of a dS
spacetime are much more complicated than those of the Minkowski
spacetime. A naive way to simplify the computation is to let the
cosmological constant $\Lambda\rightarrow 0$ directly in the field
equations. This way is equivalent to directly set
$\Lambda\rightarrow 0$ in the action. It is, however, not suitable
for the model for the following reasons. The coefficient of the
quadratic curvature term is much lager than those of the other terms
in the action. When $\Lambda\rightarrow 0$, only the quadratic
curvature term will survive in the action. As a result, in the
Einstein-like equation (see Eq. (\ref{1stEq}) in the following),
only the symmetric and trace-free part will survive when
$\Lambda\rightarrow 0$. Hence, the information of the trace part and
the antisymmetric part of the Einstein-like equation is lost when
$\Lambda\rightarrow 0$ is directly set in that equation.

In order to analyze the weak field approximation of the model, we
have presented a way to tackle the above problem \cite{WFA}, in
which the $\Lambda\rightarrow 0$ limit is set after the
Einstein-like equation is decomposed into its trace part, symmetric
trace-free part and antisymmetric part. In this way, the weak field
approximation for the static and spherically symmetric case has been
studied \cite{WFA}. It is found that there exist the solutions
describing the weak Schwarzschild fields with nonzero torsion and
the smooth connection to regular internal solutions obeying the
Newtonian gravitational law. The existence of such solutions would
determine the value of the coupling constant, which is different
from that given by Refs. \cite{Wu74,An76,Guo79,dS3}. Moreover, there
exist the solutions that could deduce the galactic rotation curves
without invoking dark matter.

The purpose of this paper is to further study the weak field
approximation of the model, taking more effects of $\Lambda$ into
consideration. Although there exist dS spacetimes with nonzero
torsion in the model \cite{dStor1,dStor2}, we assume in this paper
that the base spacetime is a torsion-free dS spacetime for
simplicity, where the base spacetime is short for the the vacuum
spacetime without perturbations from matter fields. Suppose that the
deviation from the base spacetime due to the presence of the matter
could be characterized by a parameter $s$, called the matter
parameter. The metric and torsion fields are required to be
differentiable with respect to $s$ in a neighborhood of the point
$s=0$. As was mentioned above, the computation of the deviation is
complicated, since the base spacetime is not the Minkowski
spacetime, but a dS spacetime. In this paper, the computation is
simplified by two means within two cases, respectively.

In the first case (Case A), the metric and torsion fields are
assumed to be differentiable with respect to $s$ and $\Lambda$ in a
neighborhood of the point $(s, \Lambda)=0$. The weak field
approximation is analyzed by means of expanding the
geometry and matter fields at the point $(s, \Lambda)=0$.  All the
solutions of the first-order metric and torsion fields with respect
to $s$ and $\Lambda$ are surveyed. Unfortunately, any solution will
be singular at the center ($r=0$) of the matter field, if it
describes the weak S-dS field with the smooth junction to an
internal solution.

In the second case (Case B), the metric and torsion fields are not
required to be differentiable with respect to $\Lambda$. The reason
to relax the differentiability condition is that there exist indeed
the dS spacetimes with torsion, where the torsion fields are not
differentiable with respect to $\Lambda$ at the point $\Lambda=0$
\cite{dStor1}. In this case, the computation is simplified by means
of setting $\Lambda\rightarrow 0$ in the weak field approximate
equations. Interesting enough, even if the $\Lambda\rightarrow 0$
limit is set, the effects of $\Lambda$ show up in the equations and
the corresponding solutions. Again, the smooth junction between the
weak Schwarzschild fields and regular internal solutions does not
exist in this case. The result shows that the $\Lambda\rightarrow 0$
limit of the weak field approximate equations is not equivalent to
the weak field approximation of the $\Lambda\rightarrow 0$ limit of
the field equations, even when the Einstein-like equation is
decomposed into its trace part, symmetric trace-free part and
antisymmetric part, as discussed in Ref. \cite{WFA}.

In addition, all the static and spherically symmetric solutions
which are regular at $r=0$ are presented. These regular solutions
might be used to explain the galactic rotation curves without
involving dark matter.  In GR with a cosmological constant,
although the cosmological constant would have some important effects
in the galactic-scale physics \cite{Stuchlik11}, the effects are too
small to explain the galactic rotation curves \cite{rotacurv}. On
the contrary, the possible explanation for the galactic rotation
curves is based on the torsion effects in this paper.

The paper is organized as follows. We first give a brief review to
the model of the dS gravity in section 2. In two subsections of
section 3, the static, spherically symmetric weak field
approximations of the model are calculated by two means for the two
cases, respectively. Finally, we end with some
remarks in section 4.

\section{A model of dS gravity}
A model of dS gravity has been proposed with a gauge-like action
\cite{Wu74,An76,Guo79,dS2,dS3}
\begin{eqnarray}\label{Action}
S_{G}&=&\int L_{G}=\int \kappa[-\textrm{tr}(\mathcal
{F}_{ab}\mathcal {F}^{ab})]
\nonumber\\
&=&\int\kappa[R_{abcd}R^{abcd}-\frac{4}{l^{2}}(R-\frac{6}{l^{2}})
+\frac{2}{l^{2}}S_{abc}S^{abc}]
\end{eqnarray}
in the units of $\hbar=c=1$, where $\kappa$ is a dimensionless
coupling constant, $l$ is the radius of the internal dS space and is
related to a small cosmological constant $\Lambda=3/l^2$
\cite{Wu74,An76,Guo79,dS3}, and
\begin{equation}
\mathcal {F}_{ab}=(d\mathcal {A}+\frac{1}{2}[\mathcal {A},\mathcal
{A}])_{ab}
\end{equation}
or explicitly
\begin{eqnarray}
\mathcal {F}^{A}{}_{Bab}&=&(d \mathcal {A}^{A}{}_{B})_{ab}+\mathcal
{A}^{A}{}_{Ca}\wedge\mathcal {A}^{C}{}_{Bb}
\nonumber\\
&=&\left(
\begin{array}{cc}
R_{ab}{}^{\alpha}{}_{\beta}-l^{-2}e^{\alpha}{}_{a}\wedge e_{\beta b}
&l^{-1}S^{\alpha}{}_{ab}\\
-l^{-1}S_{\beta ab}&0
\end{array}
\right)
\end{eqnarray}
is a dS algebra-valued 2-form and
\begin{equation}
\mathcal {A}^{A}{}_{Ba}=\left(
\begin{array}{cc}
\Gamma^{\alpha}{}_{\beta a}&l^{-1}e^{\alpha}{}_{a}\\
-l^{-1}e_{\beta a}&0
\end{array}
\right)
\end{equation}
is a dS algebra-valued 1-form. Here $A,B...=0,1,2,3,4$ stand for
matrix indices (internal indices) and the trace in Eq.
(\ref{Action}) is taken for those indices. In addition,
$\{e_{\alpha}{}^{a}\}$ is some local orthonormal frame field on the
spacetime manifold and $\Gamma^{\alpha}{}_{\beta a}$ is the
connection 1-form in this frame field, where $a,b...$ stand for
abstract indices \cite{{Wald},{Liang}} and
$\alpha,\beta...=0,1,2,3$. The curvature 2-form
$R_{ab}{}^{\alpha}{}_{\beta}$ and torsion 2-form $S^{\alpha}{}_{ab}$
are related to the connection 1-form $\Gamma^{\alpha}{}_{\beta a}$
as follows:
\begin{equation}
R_{ab}{}^{\alpha}{}_{\beta}=(d\Gamma^{\alpha}{}_{\beta})_{ab}+\Gamma^{\alpha}{}_{\gamma
a}\wedge\Gamma^{\gamma}{}_{\beta b},
\end{equation}
\begin{equation}
S^{\alpha}{}_{ab}=(de^{\alpha})_{ab}+\Gamma^{\alpha}{}_{\beta
a}\wedge e^{\beta}{}_{b}.
\end{equation}
Moreover,
\begin{displaymath}
R_{abc}{}^{d}=R_{ab\alpha}{}^{\beta}e^{\alpha}{}_{c}e_{\beta}{}^{d},\quad
S^{c}{}_{ab}=S^{\alpha}{}_{ab}e_{\alpha}{}^{c},
\end{displaymath}
\begin{displaymath}
R_{ab}=R_{acb}{}^{c}, \quad R=g^{ab}R_{ab},
\end{displaymath}
where the indices $\alpha,\beta$ are raised or lowered by
$\eta^{\alpha\beta}$ or $\eta_{\alpha\beta}$, respectively. The
signature is such chosen that $\eta^{00}=\eta_{00}=-1$,
$\eta^{ii}=\eta_{ii}=1$, where $i=1,2,3$. In fact, if the spacetime
is an umbilical submanifold of some (1+4)-dimensional (5d) ambient
manifold, then $\mathcal {A}_{a}$ and $\mathcal {F}_{ab}$ could be
viewed as the connection 1-form and curvature 2-form (in a
dS-Lorentz frame) of the ambient manifold restricted to the
spacetime \cite{dS3}. Here, an umbilical submanifold means a
submanifold with constant normal curvature, such as a dS spacetime
can be seen as an umbilical submanifold of the 5d Minkowski
spacetime. $\mathcal {A}_{a}$ can also be seen as the pullback of
the Cartan connection by a local section of the frame bundle
\cite{Cartan}. A principle bundle equipped with a Cartan connection
is a Cartan geometry which could be seen as a generalization of a
homogenous space, and one may refer to Ref. \cite{Cartan} for more
details.

The total action is $S=S_{M}+S_{G}$, where $S_{M}$ is the action of
the matter fields, and the field equations can be given via the
variational principle with respect to
$e^{\alpha}{}_{a},~\Gamma^{\alpha}{}_{\beta a}$:
\begin{eqnarray}\label{1stEq}
\frac{8}{l^{2}}(G_{ab}+\Lambda g_{ab})+|R|^{2}g_{ab}
-4R_{acde}R_{b}{}^{cde}+\frac{2}{l^{2}}|S|^{2}g_{ab}
\nonumber\\
-\frac{8}{l^{2}}S_{cda}S^{cd}{}_{b}
+\frac{8}{l^{2}}\nabla_{c}S_{ab}{}^{c}
+\frac{4}{l^{2}}S_{acd}T_{b}{}^{cd}+\frac{1}{\kappa}\Sigma_{ab}=0,
\end{eqnarray}
\begin{equation}\label{2ndEq}
-\frac{4}{l^{2}}T^{a}{}_{bc}-4\nabla_{d}R^{da}{}_{bc}
+2T^{a}{}_{de}R^{de}{}_{bc}
-\frac{8}{l^{2}}S_{[bc]}{}^{a}+\frac{1}{\kappa}\tau_{bc}{}^{a}=0,
\end{equation}
where
\begin{displaymath}
G_{ab}=R_{ab}-\frac{1}{2}Rg_{ab}, \quad
T^{c}{}_{ab}=S^{c}{}_{ab}+2\delta^{c}{}_{[a}S^{d}{}_{b]d},
\end{displaymath}
\begin{displaymath}
|R|^{2}=R_{abcd}R^{abcd}, \quad |S|^{2}=S_{abc}S^{abc},
\end{displaymath}
\begin{displaymath}
\Sigma_{\alpha}{}^{a}=\delta S_{M}/\delta e^{\alpha}{}_{a},\quad
\Sigma_{b}{}^{a}=\Sigma_{\alpha}{}^{a}e^{\alpha}{}_{b},
\end{displaymath}
\begin{displaymath}
\tau_{\alpha}{}^{\beta a}=\delta
S_{M}/\delta\Gamma^{\alpha}{}_{\beta a},\quad
\tau_{b}{}^{ca}=\tau_{\alpha}{}^{\beta
a}e^{\alpha}{}_{b}e_{\beta}{}^{c},
\end{displaymath}
$\nabla_{a}$ is the covariant derivative operator, which satisfies
$\nabla_a g_{bc}=0$ and is related to the connection 1-form by
\begin{displaymath}
\Gamma^{\beta}{}_{\alpha
a}=e^{\beta}{}_{b}\nabla_{a}e_{\alpha}{}^{b},
\end{displaymath}
and the variational derivatives are defined as follows: if
\begin{displaymath}
\delta S_{M}=\int(X_{\alpha}{}^{a}\delta
e^{\alpha}{}_{a}+Y_{\alpha\beta}{}^{a}\delta
\Gamma^{\alpha\beta}{}_{a}),
\end{displaymath}
then
\begin{displaymath}
\delta S_{M}/\delta e^{\alpha}{}_{a}=X_{\alpha}{}^{a},\quad \delta
S_{M}/\delta\Gamma^{\alpha\beta}{}_{a}=Y_{[\alpha\beta]}{}^{a}.
\end{displaymath}

\section{Weak field approximation}

In this section, we focus on the weak field approximation of the
model, which represents the linear deviation of the spacetime from
the base spacetime. As mentioned in Introduction, the base spacetime
is chosen to be a torsion-free dS spacetime. In a dS spacetime,
there exist Beltrami coordinate systems which play the role of
inertial coordinate systems \cite{dSSR}, where timelike/null
geodesics are all coordinate straight lines. The components of the
metric field of a dS spacetime in a Beltrami system can be written
down as follows:
\begin{equation}\label{dSmetric}
\mathring{g}_{\mu\nu}=\frac{1}{\sigma}\eta_{\mu\nu}
-\frac{1}{\sigma^{2}l^{2}}x_{\mu}x_{\nu}
=\eta_{\mu\nu}-(\eta_{\mu\nu}x^{\sigma}x_{\sigma}+x_{\mu}x_{\nu})\frac{1}{l^{2}}
+o(\frac{1}{l^{2}}),
\end{equation}
where $\eta_{\mu\nu}=\textrm{diag}(-1,1,1,1)$,
$\sigma=1+x^{\mu}x_{\mu}/l^{2}$, $x_{\mu}=\eta_{\mu\nu}x^{\nu}$ and
$\mu,\nu=0,1,2,3$. The Beltrami coordinates $x^{\mu}$ are confined
by the domain condition $\sigma>0$ \cite{dS SR}.

\subsection{Case A}

We first consider the case
where the metric and torsion fields are differentiable with respect
to $s$ and $\Lambda$ in a neighborhood of the point $(s,
\Lambda)=0$. Precisely speaking, it is assumed that
\begin{equation}\label{weakgrav}
g_{ab}=\eta_{ab}+\mathring{g}^{(1)}_{ab}+\gamma_{ab}+o(\sqrt{s^{2}+(l_{0}/l)^{4}}),\
S^{c}{}_{ab}=O(\sqrt{s^{2}+(l_{0}/l)^{4}}),
\end{equation}
\begin{equation}\label{limitgrav}
g_{ab}|_{s=0}=\mathring{g}_{ab},\quad S^{c}{}_{ab}|_{s=0}=0,
\end{equation}
\begin{equation}\label{weakmatter}
\Sigma_{ab}=O(s),\quad \tau_{ab}{}^{c}=O(s),
\end{equation}
where $s$ is a dimensionless parameter, called the matter parameter,
which describes the effects of the existence of matter fields,
$l_{0}$ is a fixed constant with dimension of length, which could be
chosen as the scale of the system under consideration,
$\gamma_{ab}\propto s$ is independent of $l$,
$\eta_{ab}=\eta_{\mu\nu}(dx^{\mu})_{a}(dx^{\nu})_{b}$,
$\mathring{g}_{ab}=\mathring{g}_{\mu\nu}(dx^{\mu})_{a}(dx^{\nu})_{b}$,
and
\begin{equation}\label{dSmetric1}
\mathring{g}^{(1)}_{ab}=-(\eta_{\mu\nu}x^{\sigma}x_{\sigma}+x_{\mu}x_{\nu})\frac{1}{l^{2}}
(dx^{\mu})_{a}(dx^{\nu})_{b},
\end{equation}
where $x^{\mu}$ are local coordinates of the spacetime, which will
come back to the Beltrami coordinates of the dS spacetime when
$s\rightarrow 0$. With respect to the assumption (\ref{weakgrav}),
the terms independent of $s$ and $l$ are regarded as the
zeroth-order terms, the terms proportional to $s$ or $1/l^{2}$ are
regarded as the first-order terms, the terms proportional to
$s^{2}$, $s/l^{2}$ or $1/l^{4}$ are regarded as the second-order
terms, and so on.  In this viewpoint, the metric field of the base
spacetime contains not only the zeroth-order term but also the
higher-order terms, as was shown in Eq. (\ref{dSmetric}). Let
$\Gamma^{c}{}_{ab}=\Gamma^{\sigma}{}_{\mu\nu}(\partial_{\sigma})^{c}(dx^{\mu})_{a}(dx^{\nu})_{b}$,
where $\Gamma^{\sigma}{}_{\mu\nu}
=(\partial_{\nu})^a(dx^{\sigma})_b\nabla_a(\partial_{\mu})^b$ is the
connection coefficient in $\{x^{\mu}\}$. $\Gamma^{c}{}_{ab}$ and the
curvature tensor have the following first-order approximate
expressions:
\begin{equation}\label{connection}
\Gamma^{c}{}_{ab}=\mathring{\Gamma}^{(1)c}{}_{ab}+\frac{1}{2}(\partial_{a}\gamma_{b}{}^{c}
+\partial_{b}\gamma_{a}{}^{c}-\partial^{c}\gamma_{ab})-K^{c}{}_{ab},
\end{equation}
\begin{equation}\label{Curvature}
R_{abc}{}^{d}=\mathring{R}^{(1)}_{abc}{}^{d}-(\partial_{c}\partial_{[a}\gamma_{b]}{}^{d}
-\partial^{d}\partial_{[a}\gamma_{b]c})+2\partial_{[a}K^{d}{}_{|c|b]},
\end{equation}
where
\begin{equation}
\mathring{\Gamma}^{(1)c}{}_{ab}=-\frac{2}{l^{2}}\delta^{\sigma}{}_{(\mu}x_{\nu)}
(\partial_{\sigma})^{c}(dx^{\mu})_{a}(dx^{\nu})_{b},
\end{equation}
\begin{equation}
\mathring{R}^{(1)}_{abc}{}^{d}=\frac{2}{l^{2}}\eta_{c[a}\delta^{d}{}_{b]}
\end{equation}
are from the dS metric field and
\begin{equation}\label{Contorsion}
K^{c}{}_{ab}=\frac{1}{2}(S^{c}{}_{ab}+S_{ab}{}^{c}+S_{ba}{}^{c})
\end{equation}
is the contorsion tensor.

The coupling constant $\kappa$ is set to be $-l^{2}/64\pi G$ in
Refs. \cite{Wu74,An76,Guo79,dS3} and $l^{2}/32\pi G$ in Ref.
\cite{WFA}. In this paper, $\kappa$ is to be determined. It has been
pointed \cite{WFA} out that when $l\rightarrow\infty$, $l^2/\kappa$
should tend to a finite value. Therefore, $1/\kappa=O(1/l^2)$, and
there does not exist first-order term in Eq. (\ref{1stEq}). The
second-order approximation of Eq. (\ref{1stEq}) and the first-order
approximation of Eq. (\ref{2ndEq}) are as follows:
\begin{equation}\label{Lambda-1stEq}
\frac{8}{l^{2}}(G_{ab}+\Lambda\eta_{ab})+|R|^{2}\eta_{ab}
-4R_{acde}R_{b}{}^{cde}+\frac{8}{l^{2}}\partial_{c}S_{ab}{}^{c}
+\frac{1}{\kappa}\Sigma_{ab}=0,
\end{equation}
\begin{equation}\label{Lambda0-2ndEq}
\partial_{d}R^{da}{}_{bc}=0.
\end{equation}
They are both equations for the first-order approximate metric and
torsion fields. The second-order approximation of Eq.
(\ref{2ndEq}) would not be considered, for the reason that it
involves the second-order approximate metric and torsion fields.

Now we restrict ourselves to the static and $O(3)$-symmetric case,
with the static spherical coordinate system
$\{T,\varrho,\theta,\varphi\}$. Let $\{\xi^{i}\}$ be related to
$\{\varrho,\theta,\varphi\}$ as follows:
\begin{equation}\label{xi-sphere}
\xi^{1}=\varrho\sin\theta\cos\varphi,\
\xi^{2}=\varrho\sin\theta\sin\varphi,\ \xi^{3}=\varrho\cos\theta.
\end{equation}
Suppose that the coordinate
systems $\{x^{\mu}\}$ and $\{T,\xi^{i}\}$ be related by:
\begin{eqnarray}\label{t-T}
&&t \equiv x^0=l\tanh(T/l)=T(1-\frac 1 3 T^2/l^2)+o(\frac{1}{l^{2}}),\\
\label{x-xi} &&
x^{i}=\xi^{i}[\cosh(T/l)\sqrt{1-\varrho^{2}/l^{2}}]^{-1}= \xi^i
(1-\frac 1 2 T^2/l^2+\frac 1 2 \varrho^2/l^2)+o(\frac{1}{l^{2}}).
\end{eqnarray}
The above relation is just the same as that of the Beltrami
coordinates and the static spherical coordinates in the dS spacetime
\cite{dSTH}. In the following discussion, the difference between $t$
and $T$, and the difference between $x^i$ and $\xi^i$ will turn out
to be negligible. In the static and $O(3)$-symmetric case, $g_{ab}$
and $S^{c}{}_{ab}$ have only these independent components in
$\{T,\varrho,\theta,\varphi\}$ \cite{torsion1}:
\begin{equation}\label{staticO3metric}
g_{TT}=g_{TT}(\varrho),\
g_{\varrho\varrho}=g_{\varrho\varrho}(\varrho),\
g_{\theta\theta}=\varrho^{2},\
g_{\varphi\varphi}=\varrho^{2}\sin^{2}\theta,
\end{equation}
\begin{equation}\label{staticO3torsion}
\left\{\begin{array}{ll}S^{T}{}_{T\varrho}=f(\varrho), \quad
S^{\theta}{}_{\varrho\theta}=S^{\varphi}{}_{\varrho\varphi}=g(\varrho),\\
S^{\varrho}{}_{T\varrho}=h(\varrho),\quad
S^{\theta}{}_{T\theta}=S^{\varphi}{}_{T\varphi}=-k(\varrho).
\end{array}\right.
\end{equation}
By Eqs. (\ref{weakgrav}),
(\ref{xi-sphere})---(\ref{staticO3torsion}), it
can be checked that the components of $\gamma_{ab}$ and the
first-order approximate $S^{c}{}_{ab}$ in $\{x^{\mu}\}$ are as
follows:
\begin{equation}\label{1stordermetric2}
\gamma_{00}=-2\phi, \quad \gamma_{0i}=0, \quad
\gamma_{ij}=(-2\psi)x_{i}x_{j}/r^{2},
\end{equation}
\begin{equation}\label{staticO3torsion2}
\left\{\begin{array}{ll}S^{0}{}_{0i}=fx_{i}/r,\quad S^{0}{}_{ij}=0,\\
S^{i}{}_{0j}=(h+k)x^{i}x_{j}/r^{2}-k\delta^{i}{}_{j},\\
S^{i}{}_{jk}=(-g/r)(\delta^{i}{}_{j}x_{k}-\delta^{i}{}_{k}x_{j}),
\end{array}\right.
\end{equation}
where $\phi$, $\psi$ are functions of
$r=\sqrt{\sum_{i=1}^{3}(x^{i})^{2}}$, and $f,\ g,\ h,\ k$ are the
first-order approximations of $f(\varrho),\ g(\varrho),\
h(\varrho),\ k(\varrho)$ with $l\rightarrow\infty$. From Eqs.
(\ref{xi-sphere}), (\ref{t-T}) and (\ref{x-xi}), when
$l\rightarrow\infty$, $\varrho\rightarrow r$. So $f,\ g,\ h,\ k$ are
functions of $r$. For the case of (\ref{staticO3torsion2}), the
contorsion tensor is related to the torsion tensor by
\begin{equation}\label{Contorsion-Torsion}
K_{abc}=S_{cba}.
\end{equation}

The equation of motion for a free spinless particle is assumed to be
the Riemannian geodesic equation:
\begin{equation}\label{particle}
U^{b}\partial_{b}U^{a}+\{^{a}_{bc}\}U^{b}U^{c}=0,
\end{equation}
where $U^{a}$ is the normalized tangent vector field of the
particle's world line and
\begin{equation}
\{^{a}_{bc}\}=\frac{1}{2}g^{ad}(\partial_{b}g_{cd}
+\partial_{c}g_{bd}-\partial_{d}g_{bc})
\end{equation}
is the Riemannian part of the connection $\Gamma^c{}_{ab}$. Suppose
that the particle's world line could be described by the functions
$x^{i}(t)$ in $\{x^{\mu}\}$. Let $u^{i}=dx^{i}/dt=
O(\sqrt{\epsilon^{2}+s+(l_{0}/l)^{2}})$, where $\epsilon$ is a
dimensionless parameter, called the low velocity parameter, which
describes the effects coming from the initial velocity $u^{i}(0)$.
Under this assumption, those terms proportional to $\epsilon,\
\sqrt{s}$ or $1/l$ are regarded as the $\frac{1}{2}$th-order terms.
Both the zeroth-order and the $\frac{1}{2}$th-order approximations
of Eq. (\ref{particle}) will give $a^{i}=0$, and the first-order
approximation of Eq. (\ref{particle}) will give
$a^{i}=-\partial^{i}\phi$, where $a^{i}=d^{2}x^{i}/dt^{2}$.
Therefore, $\phi$ may play the role of the Newtonian gravitational
potential. The equation $a^{i}=-\partial^{i}\phi$ contains the
solution which describes a circular motion with uniform speed:
\begin{equation}\label{velocity}
u^2=\left . r\frac {\partial\phi}{\partial r} \right |_{r=r(0)},
\end{equation}
where $u^2\equiv\sum^3_{i=1}(u^i)^2$, and
$r(0)=\sqrt{\sum_{i=1}^{3}[x^{i}(0)]^{2}}$ is the radius of the circular orbit.

Applying Eq. (\ref{Curvature}) to Eq. (\ref{Lambda0-2ndEq}), we have
\begin{equation}\label{Lambda0-2ndEq2}
\partial_{d}(\partial^{[d}\partial_{[c}\gamma_{b]}{}^{a]}+\partial^{[d}K_{cb}{}^{a]})=0,
\end{equation}
which gives \cite{WFA}
\begin{equation}\label{phi'+f}
\phi'+f=Cr+D/r^{2},
\end{equation}
\begin{equation}\label{k'}
h+k+rk'=0,
\end{equation}
\begin{equation}\label{Lambda0-psi}
\psi/r^{2}-g/r=B/r^{3}+A,
\end{equation}
where $A,B,C,D$ are constants of integration.

From Eqs. (\ref{Curvature}), (\ref{1stordermetric2}) and
(\ref{staticO3torsion2}), the leading components of $R_{abcd}$ in
$\{x^{\mu}\}$ could be attained as follows:
\begin{equation}
\left\{\begin{array}{ll}R_{0i0j}=-\delta_{ij}/l^{2}+(\phi''+f')x_{i}x_{j}/r^{2}+(\phi'+f)(\delta_{ij}r^{2}-x_{i}x_{j})/r^{3},\\
R_{0ijk}=0,\\
R_{ijk0}=(2/r^{2})(h+k+rk')x_{[i}\delta_{j]k},\\
R_{ijkl}=2\delta_{i[k}\delta_{l]j}/l^{2}+(\psi/r^{2}-g/r)'(4/r)x_{[i}\delta_{j][k}x_{l]}\\
\qquad \quad -4(\psi/r^{2}-g/r)\delta_{i[k}\delta_{l]j}.
\end{array}\right.
\end{equation}
Substituting Eqs. (\ref{phi'+f}), (\ref{k'}) and (\ref{Lambda0-psi})
into the above equation, we obtain
\begin{equation}\label{reducedRiemann}
\left\{\begin{array}{ll}R_{0i0j}=-\delta_{ij}/l^{2}+(Cr+D/r^{2})'x_{i}x_{j}/r^{2}+(Cr+D/r^{2})(\delta_{ij}r^{2}-x_{i}x_{j})/r^{3},\\
R_{0ijk}=0,\\
R_{ijk0}=0,\\
R_{ijkl}=2\delta_{i[k}\delta_{l]j}/l^{2}+(B/r^{3}+A)'(4/r)x_{[i}\delta_{j][k}x_{l]}
-4(B/r^{3}+A)\delta_{i[k}\delta_{l]j}.
\end{array}\right.
\end{equation}
Then
\begin{equation}\label{sym-tracefree}
\begin{array}{ll}
|R|^{2}\eta_{00}-4R_{0cde}R_{0}{}^{cde}=(-24/l^{2})(C-2A)+12(C^{2}-4A^{2})+24(D^{2}-B^{2})/r^{6},\\
|R|^{2}\eta_{0i}-4R_{0cde}R_{i}{}^{cde}=0,\\
|R|^{2}\eta_{ij}-4R_{icde}R_{j}{}^{cde}=(-8/l^{2})(C-2A-2B/r^{3}-2D/r^{3})\delta_{ij}\\
\qquad \qquad \quad +(-48/l^{2})(B/r^{3}+D/r^{3})x_{i}x_{j}/r^{2}\\
\qquad \qquad \quad +(4C^{2}-16A^{2}-16CD/r^{3}-32AB/r^{3}+16D^{2}/r^{6}-16B^{2}/r^{6})\delta_{ij}\\
\qquad \qquad \quad +(48CD/r^{3}+96AB/r^{3}
-24D^{2}/r^{6}+24B^{2}/r^{6})x_{i}x_{j}/r^{2},
\end{array}
\end{equation}
\begin{equation}\label{Einstein+Lambda}
\begin{array}{lll}
G_{00}+\Lambda\eta_{00}&=& -6A,\\
G_{0i}+\Lambda\eta_{0i}&=& G_{i0}+\Lambda\eta_{i0}=0,\\
G_{ij}+\Lambda\eta_{ij}&=&(2C+2A-D/r^{3}-B/r^{3})\delta_{ij}
+[3(B+D)/r^{3}]x_{i}x_{j}/r^{2}.
\end{array}
\end{equation}

From Eq. (\ref{staticO3torsion2}), the leading components of
$\partial_{c}S_{ab}{}^{c}$ could be given as follows:
\begin{equation}
\begin{array}{ll}\partial_{c}S_{00}{}^{c}=-f'-2f/r,\\
\partial_{c}S_{0i}{}^{c}=0,\\
\partial_{c}S_{i0}{}^{c}=x_{i}[h'/r+2(h+k)/r^{2}],\\
\partial_{c}S_{ij}{}^{c}=[-2(g/r)-r(g/r)']\delta_{ij}+(g/r)'x_{i}x_{j}/r.
\end{array}
\end{equation}
Substituting Eqs. (\ref{phi'+f}) and (\ref{Lambda0-psi}) into
the above equation, we have
\begin{equation}\label{derivativetorsion}
\begin{array}{ll}\partial_{c}S_{00}{}^{c}=\triangle\phi-3C,\\
\partial_{c}S_{0i}{}^{c}=0,\\
\partial_{c}S_{i0}{}^{c}=x_{i}[h'/r+2(h+k)/r^{2}],\\
\partial_{c}S_{ij}{}^{c}=(-B/r^{3}+2A-\psi'/r)\delta_{ij}+[r(\psi/r^{2})'+3B/r^{3}]x_{i}x_{j}/r^{2},
\end{array}
\end{equation}
where $\triangle\phi=\phi''+(2/r)\phi'$.

With the help of Eqs. (\ref{sym-tracefree}), (\ref{Einstein+Lambda})
and (\ref{derivativetorsion}), Eq. (\ref{Lambda-1stEq}) can
be solved. Note that from Eqs. (\ref{weakgrav})
and (\ref{limitgrav}), $S^{c}{}_{ab}\propto s$ in the first-order
approximation, and so all of the constants $A,\ B,\ C$ and $D$ are
proportional to $s$. Eq. (\ref{Lambda-1stEq}) splits into two
equations, one for those terms proportional to $s^{2}$ and the other
one for those terms proportional to $s/l^{2}$, while those terms
proportional to $1/l^4$ add up to zero. The $s^{2}$ equation yields
the following relations:
\begin{equation}\label{CDAB}
C=\pm 2A,\ B=\pm D,\ CD+2AB=0.
\end{equation}
Suppose that $\Sigma_{ab}=\rho (dt)_{a}(dt)_{b}$. Then the $00$
component of the $s/l^{2}$ equation is
\begin{equation}\label{1stEq00}
\triangle\phi-6C+(l^{2}/8\kappa)\rho=0,
\end{equation}
the $0i$ component is an identity and the $i0$ component gives
\begin{displaymath}
h'/r+2(h+k)/r^{2}=0
\end{displaymath}
with the solution
\begin{equation}\label{h-k}
h=2k+C_{1},\quad h+k=C_{2}/r^{3},
\end{equation}
where Eq. (\ref{k'}) has been used. Finally, the $ij$ component
gives
\begin{displaymath}
(C+6A+D/r^{3}-\psi'/r)\delta_{ij}+[r(\psi/r^{2})'-3D/r^{3}]x_{i}x_{j}/r^{2}=0
\end{displaymath}
with the solution
\begin{equation}\label{psi}
\psi=\frac{1}{2}(C+6A)r^{2}-D/r.
\end{equation}

For the S-dS solution, no matter with or without torsion,
there should be $\phi=\psi=-GM/r$ and so $D=GM$. If it has a smooth
connection to an internal solution, then the integration constant
$D=GM$ should also hold for that internal solution. As a result, the
internal solution will be singular at $r=0$.  Such a singular
solution may represent a strong field near $r = 0$. It should be
stressed that it is the solution of the first-order gravitational
field under the assumptions (\ref{weakgrav})--(\ref{weakmatter}).
The solution shows that if the interior is smoothly connected to the
S-dS exterior, its first-order term should be singular at $r = 0$.
Therefore, the interior in the full theory should be singular at
$r=0$, too.

For those solutions which are regular at $r=0$, we should let
$B=D=0$ and $C_{2}=0$ from Eqs. (\ref{phi'+f}), (\ref{Lambda0-psi})
and (\ref{h-k}). Then
\begin{equation}\label{regularpsi}
\psi=\frac{1}{2}(C+6A)r^{2},
\end{equation}
\begin{equation}\label{regulartorsion}
f=Cr-\phi',\quad g=\frac{1}{2}(C+4A)r,\quad h=-k=\frac{1}{3}C_{1},
\end{equation}
and $\phi$ is still given by Eq. (\ref{1stEq00}). Generally, the
torsion tensor can be decomposed \cite{{torsion2},{Hehl95}} into
three irreducible parts with respect to the Lorentz group: the
tensor part, trace-vector part, and the axial vector part. For
static and $O(3)$-symmetric torsion, the axial vector part vanishes
automatically, the tensor part satisfies $f=2g$, $h=2k$ and the
trace-vector part satisfies $f=-g$, $h=-k$. In the above solutions,
if the torsion tensor only contains the trace-vector part, then
\begin{displaymath}
\phi'=\frac{1}{2}(3C+4A)r,\quad \triangle\phi=\frac{3}{2}(3C+4A).
\end{displaymath}
If the torsion tensor only contains the tensor part, then
\begin{displaymath}
\phi'=-4Ar,\quad \triangle\phi=-12A.
\end{displaymath}
Therefore the above two cases can only describe matter fields with
constant density.

As a special choice, we may let $C=A=0$ and $C_{1}=0$, then
\begin{equation}\label{special}
\triangle\phi+(l^{2}/8\kappa)\rho=0,\quad \psi=0,\quad
f=-\phi',\quad g=h=k=0.
\end{equation}
In this solution, the curvature tensor is just the same as that of
the dS spacetime, and the torsion tensor has the property that
both the tensor and trace-vector parts
have nonzero values unless $\rho=0$. A
comparison between Eq. (\ref{special}) and $\triangle\phi=4\pi
G\rho$ would fix the value of the coupling constant:
\begin{equation}\label{kappa}
\kappa=-l^{2}/32\pi G,
\end{equation}
which is different from those of Refs.
\cite{Wu74,An76,Guo79,dS3,WFA}. The choice in Refs.
\cite{Wu74,An76,Guo79,dS3} is to guarantee the ratio of the
coefficient of $G_{ab}$ to that of $\Sigma_{ab}$ in Eq.
(\ref{1stEq}) be $1:(-8\pi G)$, just like the case in GR. But it has
been pointed \cite{WFA} out that the Einstein term in our model
plays a completely different role from that of GR. By Eq.
(\ref{Einstein+Lambda}), the 00 component of the Einstein term with
the $\Lambda$ term only contributes a constant term, which would be
canceled by the quadratic curvature term and does not show up in Eq.
(\ref{1stEq00}). In the $\Lambda\rightarrow 0$ limit in Ref.
\cite{WFA}, the choice of the coupling constant is to guarantee the
existence of the smooth junction between the weak Schwarzschild
fields and regular internal solutions obeying the Newtonian
gravitational law. But that smooth junction has no correspondence in
the present case with $\Lambda\nrightarrow 0$.

Generally, the regular solutions are given by Eqs.
(\ref{regularpsi}), (\ref{regulartorsion}) and
\begin{equation}\label{1stEq004pi}
\triangle\phi-6C=4\pi G\rho,
\end{equation}
where Eq. (\ref{kappa}) has been used. The integration of Eq.
(\ref{1stEq004pi}) gives
\begin{equation}\label{phiprime}
\phi'= Gm(r)/r^2+2Cr,
\end{equation}
where
\[
m(r)=4\pi\int_0^r \rho r^2 dr.
\]
Substituting Eq. (\ref{phiprime}) into Eq. (\ref{velocity}) results in
\begin{equation}\label{velocity1}
u^2(r)=Gm(r)/r+2Cr^2,
\end{equation}
which describes the speed of a free particle revolving around the
center of the gravitational field. Suppose that the coordinate
radius of the matter surface is $R_s$, and denote $m(R_s)$ by $M$.
In the region with $r> R_s$, Eq. (\ref{velocity1}) can be viewed as
the first-order term of
\begin{equation}\label{velocity2}
v^2(r)=\frac{GM/r +2C r^2}{1-2GM/r+2C r^2},
\end{equation}
which is the square of the rotation speed in GR with a cosmological
constant $\Lambda_0$, relative to an inertial observer which is
instantaneously at rest in the static frame of reference, if
$\Lambda_0=-6C$. Eq. (\ref{velocity2}) with $\Lambda_0
=-10^{-48}$m$^{-2}$ ($C\approx 10^{-49}$m$^{-2}$) has been used to
fit the flat rotation curves for the galaxies NGC 224, 2841, 2903,
etc without involving dark matter \cite{rotacurv}. Hence, Eq.
(\ref{velocity1}), as the first order approximation of Eq.
(\ref{velocity2}), might also be used to explain the flat rotation
curves in the same way with the same value for $C$. It is worth
mentioning that the constant $C$ represents the torsion effects
here. It may get rid of the inconsistency of the value and the sign
of the cosmological constant with cosmological observations in Ref.
\cite{rotacurv}. In fact, Eqs. (\ref{CDAB}) and
(\ref{regulartorsion}) show that $C$ is related to the torsion
component $g$ by
\begin{equation}
C=\frac {2g}{(1\pm 2)r}.
\end{equation}
Finally, it should be remarked that Eq. (\ref{velocity2}) would no
longer be valid in our model. The higher-order behavior of the
rotation speed function will be different from that in GR.

\subsection{Case B}
Now, we turn to the general case where the metric field and torsion
field are not required to be differentiable with respect to
$\Lambda$. The assumption on matter fields is still given by Eq.
(\ref{weakmatter}), and the geometry is assumed to be
\begin{equation}\label{weakgrav2}
g_{ab}=\mathring{g}_{ab}+\gamma_{ab}+o(s),\quad S^{c}{}_{ab}=O(s),
\end{equation}
where $\gamma_{ab},\ o(s)$ and $O(s)$ may depend on $l$. From Eqs.
(\ref{dSmetric}), (\ref{dSmetric1}) and (\ref{weakgrav2}), we have
\begin{equation}
g_{ab}=\eta_{ab}+\mathring{g}^{(1)}_{ab}+\gamma_{ab}+o(\frac{1}{l^{2}})+o(s),
\end{equation}
where $o(1/l^{2})$ is independent of $s$. The expressions for the
connection and curvature are as follows:
\begin{equation}
\Gamma^{c}{}_{ab}=\mathring{\Gamma}^{(1)c}{}_{ab}+\frac{1}{2}(\partial_{a}\gamma_{b}{}^{c}
+\partial_{b}\gamma_{a}{}^{c}-\partial^{c}\gamma_{ab})-K^{c}{}_{ab}
+o(\frac{1}{l^{2}})+O(\frac{1}{l^{2}})O(s)+o(s),
\end{equation}
\begin{equation}\label{curvatrue2}
R_{abc}{}^{d}=\mathring{R}^{(1)}_{abc}{}^{d}-(\partial_{c}\partial_{[a}\gamma_{b]}{}^{d}
-\partial^{d}\partial_{[a}\gamma_{b]c})+2\partial_{[a}K^{d}{}_{|c|b]}
+o(\frac{1}{l^{2}})+O(\frac{1}{l^{2}})O(s)+o(s).
\end{equation}
For simplicity, we only consider the $\Lambda\rightarrow 0$ limit of
the weak field approximate equations. With respect to the assumption
(\ref{weakgrav2}), those terms independent of $s$ are regarded as
the zeroth-order terms, those terms proportional to $s$ are regarded
as the first-order terms, those terms proportional to $s^{2}$ are
regarded as the second-order terms, and so on. Then the
$\Lambda\rightarrow 0$ limit of the second-order approximation of
Eq. (\ref{1stEq}) and the first-order approximation of Eq.
(\ref{2ndEq}) are:
\begin{equation}\label{1st2ndEqLambda0}
|R|^{2}\eta_{ab}-4R_{acde}R_{b}{}^{cde}=0,\quad
\partial_{d}R^{da}{}_{bc}=0
\end{equation}
with the solutions given by Eqs. (\ref{phi'+f}), (\ref{k'}),
(\ref{Lambda0-psi}) and (\ref{CDAB}). The $\Lambda\rightarrow 0$
limit of the first-order approximation of Eq. (\ref{1stEq}) is
\begin{equation}\label{1stEqLambda0}
8G_{ab}+l^{2}(|R|^{2}\eta_{ab}
-4R_{acde}R_{b}{}^{cde})|_{l\rightarrow\infty}+8\partial_{c}S_{ab}{}^{c}
+(l^{2}/\kappa)|_{l\rightarrow\infty}\Sigma_{ab}=0.
\end{equation}
Obviously, the terms $o(\frac{1}{l^{2}})+O(\frac{1}{l^{2}})O(s)$ in
Eq. (\ref{curvatrue2}) do not contribute to Eq.
(\ref{1stEqLambda0}). The solutions of Eq. (\ref{1stEqLambda0})
would then be given by Eqs. (\ref{1stEq00}), (\ref{h-k}) and
(\ref{psi}). Therefore, the smooth junction between the weak
Schwarzschild fields and regular internal solutions does not exist.
It shows that the $\Lambda\rightarrow 0$ limit of the weak field
approximation is not equivalent to the weak field approximation of
the $\Lambda\rightarrow 0$ limit discussed in Ref. \cite{WFA}. In
addition, all solutions of Eqs. (\ref{1st2ndEqLambda0}) and
(\ref{1stEqLambda0}), which are regular in the finite $r$ region,
are given by Eqs. (\ref{1stEq00}), (\ref{regularpsi}) and
(\ref{regulartorsion}). They might be used to explain the galactic
rotation curves, as discussed in the last subsection.

\section{Remarks}
The weak field approximation of the dS gravity model is analyzed in
the static and spherically symmetric case, under the assumption that
the base spacetime is a torsion-free dS spacetime. Concretely, we
study the weak field approximation by two means within two cases,
respectively. In Case A, the metric and torsion fields are
required to be differentiable with respect to the matter
parameter $s$ and the cosmological constant $\Lambda$ in a vicinity of
the point $(s,\Lambda)=0$. We expand the field
equations with respect to $s$ and $\Lambda$, and calculate the
first-order approximate metric and torsion fields. In Case B, the
metric and torsion fields are not required to be differentiable with
respect to $\Lambda$. We expand the field equations with respect to
$s$, and solve the $\Lambda\rightarrow 0$ limit of the weak field
approximate equations. Unfortunately, for both cases, we reach the
conclusion that the S-dS solutions could not
be smoothly connected to regular internal solutions.

For those solutions which are regular everywhere in the finite $r$
region, we find that due to the torsion effects, they might be used
to deduce the galactic rotation curves without involving dark
matter. In addition, in the regular solutions, both the tensor and
trace-vector parts of the torsion field should be nonzero in a
generic case with inhomogeneous matter density. It shows that the
torsion tensor plays an important role in this model, as was pointed
out by Ref. \cite{WFA}. On the other hand, the curvature tensor has
a special form, for example, given by Eq. (\ref{reducedRiemann}) in
Case A. As a result, the Einstein term in this model plays a
completely different role from that of GR. The choice of the
coupling constant $\kappa=-l^{2}/64\pi G$ \cite{Wu74,An76,Guo79,dS3}
may be problematic, which is due to the comparison between GR and
the dS gravity model. In fact, if and only if $\kappa=-l^{2}/32\pi
G$ is set, there exist the solutions which are in accordance with
the Newtonian gravitational law in the two cases of this paper, with
$\Lambda\nrightarrow 0$ and $\Lambda\rightarrow 0$, respectively. In
the $\Lambda\rightarrow 0$ limit discussed in Ref. \cite{WFA},
$\kappa$ is set to be $l^{2}/32\pi G$, which is to guarantee the
existence of the smooth junction between the weak Schwarzschild
fields and regular internal solutions obeying the Newtonian
gravitational law. But the limit cannot be attained from the two
cases in the present paper.

Similar to the Schwarzschild solution in GR, the S-dS solution in
the dS gravity model plays an important role in the explanation of
the solar-system-scale observations. For the theory consistent with
the observations within the solar system, the S-dS solution should
be able to link to the regular internal solution smoothly. From the
results of this paper, if there exist the S-dS solutions with the
smooth junction to regular internal solutions, they would not
satisfy Eqs. (\ref{weakmatter}) and (\ref{weakgrav2}) or would be
singular in the $\Lambda\rightarrow 0$ limit. Therefore, the future
studies may be based on this fact. Especially, since the torsion
tensor plays an important role, one may try to see what would happen
if the base spacetime is a dS spacetime with torsion, the examples
of which have been obtained in Refs. \cite{dStor1,dStor2}. If there
does not exist the torsional S-dS solution with the smooth junction
to a regular internal solution, the remaining way is to study higher
order approximation of the gravitational field equations as well as
the particle's motion, which might be the only possible way to make
the model consistent with the observations within the solar system.

\phantomsection \addcontentsline{toc}{section}{Acknowledgments}
\section*{Acknowledgments}
We would like to thank Prof. Zhe Chang and thank late Prof. Han-Ying
Guo for useful discussions. One of us (Lu) would like to thank
Profs. Zhi-Bing Li and Xi-Ping Zhu for their help and guidance. This
work is supported by National Natural Science Foundation of
China under Grant Nos. 10975141, 10831008, 11275207,
and the oriented projects of CAS under Grant No. KJCX2-EW-W01.

\phantomsection 
\end{document}